# Dielectric breakdown and sub-wavelength patterning of monolayer hexagonal boron nitride using femtosecond pulses

Sabeeh Irfan Ahmad[1], Emmanuel Sarpong[1], Arpit Dave[1], Hsin-Yu Yao[2], Joel M. Solomon[1], Jing-Kai Jiang[3], Chih-Wei Luo[3], Wen-Hao Chang[3,4] and Tsing-Hua Her[1]

[1]Department of Physics and Optical Science, The University of North Carolina at Charlotte, 9201 University City Boulevard, Charlotte NC 28223
[2]Department of Physics, National Chung Cheng University, No.168, Sec. 1, University Rd., Minhsiung, Chiayi 621301, Taiwan.
[3]Department of Electrophysics, National Yang Ming Chiao Tung University, Hsinchu 30010, Taiwan
[4]Research Center for Applied Sciences, Academia Sinica, Taipei 11529, Taiwan

Email: ther@uncc.edu,

Hexagonal boron nitride (hBN) has emerged as a promising two-dimensional (2D) material for many applications in photonics. Although its linear and nonlinear optical properties have been extensively studied, its interaction with high-intensity laser pulses, which is important for high-harmonic generation, fabricating quantum emitters, and maskless patterning of hBN, has not been investigated. Here we report the first study of dielectric breakdown in hBN monolayers induced by single femtosecond laser pulses. We show that hBN has the highest breakdown threshold among all existing 2D materials. This enables us to observe clearly for the first time a linear dependence of breakdown threshold on the bandgap energy for 2D materials, demonstrating such a linear dependency is a universal scaling law independent of the dimensionality. We also observe counter-intuitively that hBN, which has a larger bandgap and mechanical strength than quartz, has a lower breakdown threshold. This implies carrier generation in hBN is much more efficient. Furthermore, we demonstrate the clean removal of hBN without damage to the surrounding hBN film or the substrate, indicating that hBN is optically very robust. The ablated features are shown to possess very small edge roughness, which is attributed to its ultrahigh fracture toughness. Finally, we demonstrate femtosecond laser patterning of hBN with sub-wavelength resolution, including an isolated stripe width of 200 nm. Our work advances the knowledge of light-hBN interaction in the strong field regime and firmly establishes femtosecond lasers as novel and promising tools for one-step deterministic patterning of hBN monolayers.

# Introduction

Hexagonal boron nitride (hBN) has emerged as a promising two-dimensional (2D) material for many applications in electronics and photonics due to its large bandgap, mechanical flexibility and breaking strength [1], high thermal conductivity [1], and chemical stability [2]. Its linear [3] and nonlinear optical properties in the weak limit [4, 5] have been extensively studied. To date, little is known about hBN's interaction with the strong field in high-intensity laser pulses. Such knowledge is very useful for a variety of applications. For example, high-harmonic generation in hBN has been theoretically predicted but not yet demonstrated experimentally [6]. Knowledge of material degradation and thresholds of optical dielectric breakdown (ODB) induced by high-intensity ultrashort pulses would be particularly useful for this future work. As another example, spin defects in hBN have been extensively studied as quantum emitters due to their brightness and spin-dependent emission at room temperature [7]. Engineering deterministic defects in hBN induced by femtosecond laser pulses has been actively pursued, revealing that spin defects were only found around the edge of ablated features [7, 8]. Understanding the physics of ODB in hBN therefore constitutes the first step towards engineering deterministic defects in hBN. Finally, many electrical and photonic applications of hBN require pattering. For example, a one-dimensional grating made of thin hBN layers is a mid-infrared hyperbolic metasurface that supports strongly volume-confined phonon polaritons with a diverging large photon density of states [9]. For another example, various micro-optical components including photonic crystal cavities, micro-ring resonators, and waveguides with grating couplers have been demonstrated in hBN, which are essential building blocks for a monolithic quantum photonics platform integrated with spin defects as single photon sources [10]. Numerous methods have been demonstrated for patterning hBN, but each has drawbacks. Lithography based on photon, electron-beam (e-beam), or atomic force microscopy (AFM) tends to leave photoresist traces behind that can degrade device performance [11]. Direct etching based on ion beam or e-beam tends to leave damage in the surroundings of the patterned regions due to secondary electrons randomly backscattered from the substrate [12]. Furthermore, processes involving e-beam or ion beams are slow and require complicated vacuum environment [13]. Compared to these, laser patterning can be advantageous, as it avoids the use of resists, forgoes the need of high vacuum, and is a high-speed technique for prototyping. Previously Stohr et al. demonstrated the fabrication of holes of ~ 85 nm in diameter and stripes of ~ 20 nm in width in graphene using a 10-ps spatially structured beam via thermally induced oxidative burning of graphene [14]. Solomon et al. demonstrated stripes of ~ 250 nm in width in $MoS_2$ monolayers by femtosecond laser ablation [15]. To date, although femtosecond laser ablation of hBN has been reported [7, 8], the physics of dielectric breakdown in hBN is not known, and laser patterning of hBN has not been demonstrated.

Here, we report the first systematic study of dielectric breakdown of hBN monolayers induced by a single femtosecond laser pulse. We show that hBN has the highest breakdown threshold among all 2D materials. This enables us to reveal clearly for the first time a linear dependence of breakdown threshold on the bandgap energy of 2D materials, indicating that such a linear dependency is a universal scaling law, independent of the dimensionality. Moreover, we provide indirect evidence for strongly enhanced carrier generation in hBN compared to bulk supporting substrates with similar bandgap, which is attributed to enhanced nonlinear ionization and/or enhanced avalanche ionization in 2D materials. Furthermore, we demonstrate a clean removal of hBN while leaving the surrounding hBN film and the substrate intact, indicating that hBN is optically very robust. The ablated features in hBN were revealed to have very small edge roughness due to its ultrahigh fracture toughness. Finally, we demonstrate femtosecond laser patterning of hBN with sub-micron resolution. Our work clearly positions femtosecond laser ablation as a promising approach for patterning hBN monolayers.

## Materials and methods

### Sample preparation

The hBN monolayers were grown on 500-nm-thick Cu(111) on c-plan $Al_2O_3$ substrates by chemical vapor deposition (CVD) using ammonia borane (97%) as the precursor. The as-grown monolayer hBN film was detached from the Cu(111)/ $Al_2O_3$ substrate by electrochemical delamination using a poly(methyl methacrylate) (PMMA) film and a thermal release tape (TRT) as the supporting layer. After detachment, the TRT/PMMA/hBN stacked film was placed on the target substrate. The TRT and PMMA film were finally removed by baking and hot acetone, leaving behind a monolayer hBN film on the target substrate. PMMA residue was further removed using rapid thermal annealing at 400°C for 1 min under a 6-torr forming gas (5% $H_2$ and 95% nitrogen). Further details of the growth and transfer can be found in [16].

In the literature, while hBN monolayer is still arguably a direct or indirect bandgap material [17], the reported direct bandgap of hBN monolayers at K point ranges from 7.2 (minimum) - 8.2 eV (maximum) [3, 17, 18]. In this work, we quote its bandgap as 7.6 ± 0.3 eV. To investigate the optical dielectric breakdown (ODB) of hBN monolayers, an optically robust substrate is needed, which disqualifies 90 nm $SiO_2$-Si substrates [8]. Instead, our experiments were carried out on transparent substrates including single-crystal $Al_2O_3$ and fused quartz. Their UV-VIS absorption spectra were taken using a Shimadzu UV2600 spectrophotometer (see section S1), from which we determined the bandgap of fused quartz to be ~ 5.8 eV. $Al_2O_3$ remained transparent within instrument's detection limit. As the literature reported value ranges between 8 (minimum) and 9.4 eV (maximum) [19, 20], in this work, we quote its bandgap as 8.8 ± 0.4 eV.

After the laser exposure, attempts were made to remove PMMA residue by washing the samples either in an acetone vapor bath at 140 °C for 30 mins or fully immersing in the acetone at 80 °C for 30 mins with magnetic agitation.

## Laser ablation and patterning setup

The laser employed in this experiment is a Coherent RegA 9000, producing 800-nm 160-fs pulses with a pulse energy stability of ~ 0.5%. For the single-shot experiment, the laser was operated at 300 Hz and a single pulse was selected using a mechanical shutter. The laser was focused by either a 0.26-NA (Mitutoyo NIR objective, 10×) or a 0.9-NA (Leitz-Wetzlar NPL 100×) objective. The sample was mounted to a 3-axis translation stage (Aerotech ANT-50L) to position at the laser focus and for lateral position control. A circular ND filter wheel was used to select the pulse energy. For line patterning, the laser was operated at 100 kHz and focused through the 0.9-NA objective, and the translation speed was 100 $\mu m/s$.

## Microscopy

For graphene and $MoS_2$ the ablated features were imaged using an Olympus BX-51 optical microscope (OM). For hBN, the ablated holes were not visible under optical microscopes; instead, AFM (Veeco Nanoscope IIIa) operated in tapping mode with a tip diameter of 10 nm was used for imaging. The open-source software Gwyddion was used for AFM image processing.

## Raman

For material analysis, Raman spectra were collected using a home-built epi-luminescent microscope with a 0.9-NA (Olympus 100×) objective. A dichroic filter (Long pass 532 nm, Iridian ZX827) was used to separate the 532 nm continuous-wave excitation (Lasos GLK 3250 T01) from the Raman signal, which was then coupled into a 200 $\mu m$ fiber and detected using a Czerny-Turner imaging spectrometer (Horiba iHR-550) with a liquid nitrogen cooled CCD (Symphony). In the literature, the vast majority of Raman studies of hBN monolayers were conducted on 90 nm $SiO_2$/Si substrates where the etalon effect enhanced the signal strength [15]. On transparent substrates which lack the etalon enhancement, the Raman signal strength is extremely weak, which prompted us to use a high laser power (~ 5 mW) and long integration time (~ 20 min). To further increase the signal-to-noise ratio, a larger focused beam radius ($\omega_o$ = 1.16 µm) was used, which compromises the spatial resolution. It is worth pointing out that Raman measurements can only be conducted on quartz, as $Al_2O_3$ has a Raman peak at 1356 cm$^{-1}$ that interferes with that of the hBN monolayers (see section S2).

## Fidelity analysis of laser ablation

To quantify the fidelity of laser ablation, we calculate the percentage area mismatch between the ablated features and the scaled focused laser beam profile. The former were imaged via either OM or AFM, while the latter was imaged using a home-built optical microscope equipped with a 0.8 NA 100× objective and a CCD (WinCamD UCD23). The laser beam profile was scaled to match the ablated features, from which the area difference (in terms of pixel counts) between these two were computed. The above difference is normalized to the ablated area yielding the percentage area mismatch. For more details, refer to section S3.

# Results and discussion

## Breakdown threshold and the etalon effect

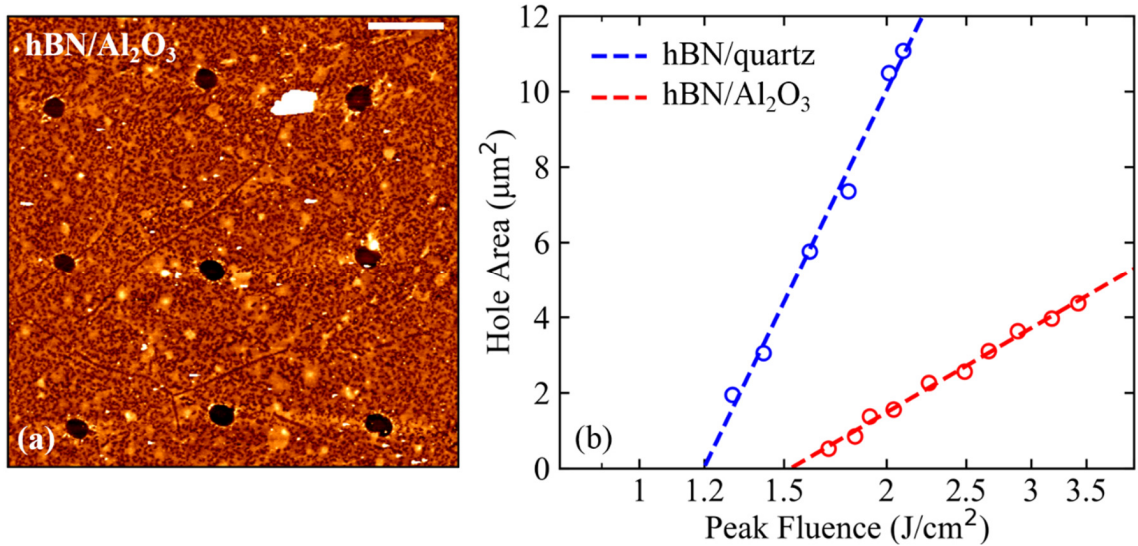

Figure 1: (a) An AFM image of ablated features in a hBN monolayer on $Al_2O_3$ with increasing laser fluence in columns from left to right. The focused laser beam radius is 1.87 μm. The scale bar is 5 μm. (b) A linear-log plot of ablated hole area vs. peak laser fluence for hBN monolayer on $Al_2O_3$ (red trace) and quartz (blue trace). The focused laser spot sizes are different for these two experiments.

Figure 1(a) shows an AFM image of an array of ablated holes in a hBN monolayer on $Al_2O_3$ with fluences 1.82 J/cm² (left), 1.93 J/cm² (middle), and 2.07 J/cm² (right). The scaling of the hole area with the laser fluence is evident. This procedure was repeated for different fluences and supporting substrates ($Al_2O_3$ and quartz). The resulting traces (hole area $A$ vs. peak fluence $F$) are displayed in Figure 1(b) and fitted to Liu's equation $A = (\pi/2)\omega_o^2 \ln(F/F_{th})$ [21] where $\omega_o$ is Gaussian beam radius ($1/e^2$-intensity) and $F_{th}$ is the ablation threshold. The extracted $\omega_o$ and $F_{th}$ are respectively 3.53 µm and 1.2 J/cm² for quartz, and 1.87 µm and 1.53 J/cm² for Al₂O₃. The difference in laser spot size is due to different collimated beam radii before the focusing objective. The difference in $F_{th}$ can be explained by the etalon effect. Previously we have shown that $F_{th}$ varies with the supporting substrates due to the etalon effect [15], where the internal electric field $E'$ inside the 2D materials scales with the incident electric field $E$ according to $E' = \eta E$ with $\eta = 2/(1+n_s)$ being the field enhancement factor and $n_s$ being the substrate's refractive index. We further showed that $F_{th}\eta^2$ is a substrate-independent quantity, from which we define an intrinsic ODB fluence $F_{th}^{intr}$ which corresponds to the incident $F_{th}$ for a free-standing 2D material where $E' = E$. In our experiment, $\eta_{SiO_2}^2 = 0.66$ and $\eta_{Al_2O_3}^2 = 0.53$ at 800 nm, which render $F_{th}^{intr}$ of hBN to be 0.79 and 0.81 J/cm² for quartz and Al₂O₃ substrates, respectively. Our result is therefore consistent with the etalon effect, and we obtain $F_{th}^{intr}$ for a hBN monolayer to be ~ 0.8 J/cm² for 160-fs pulses. We emphasize that it is essential to quote $F_{th}^{intr}$ when comparing $F_{th}$ of dissimilar 2D materials on different substrates as well as when comparing with theories which usually invoke the internal electrical field for light-matter interaction. We note that the etalon effect was previously invoked in the femtosecond ODB of dielectric thin films of hundreds of nanometers thick [22]; our previous and current work prove that etalon effect holds true even for atomic layers.

## Morphology and patterning fidelity

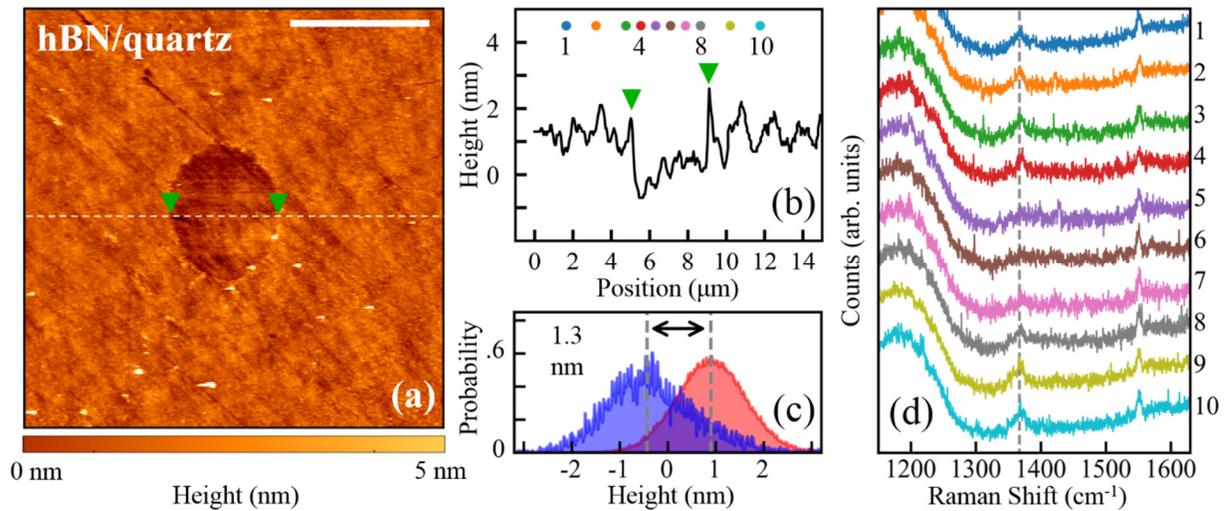

Figure 2: (a) Ablated hole in hBN on fused quartz with 1.3 J/cm² incident fluence. The focused laser beam radii (1/e²-intensity) is 6.32 $\mu m$. The white particles of varying shapes and sizes are PMMA residue from the transfer process. Scale bar is 5 µm. (b) Cross-sectional line profile of a 1.2 nm thick strip of (a) represented by the dotted line. Green arrows indicate the edge of the hole. (c) Histogram of height data of panel (a) inside the hole (blue data) and region outside the hole (red data). (d) Raman signals collected across (a). The signals correspond to colored markers in (b).

Figure 2(a) shows an ablated hole on fused quartz obtained with a fluence of 1.3 J/cm$^2$, whose cross-sectional height profile (averaged over a horizontal strip 1.2 nm wide) along the dashed line is displayed in Figure 2(b) with green arrows indicating the edges of the hole. Figure 2(c) shows histograms of AFM height data within the hole (blue trace) and of the pristine film (red trace). The difference in the peak positions indicates a film height of ~ 1.3 nm, which is consistent with the literature value of 0.3 – 2.8 nm for hBN monolayers [1, 8]. The width of the histogram for the hole is 1.95 nm, which is slightly larger than 1.76 nm of the pristine film, suggesting that the hBN film acts like a carpet to hide the surface roughness of the underlying bare substrate. The surface roughness inside the hole (Figure 2(c)) is the same as that of the bare substrate (see section S4), indicating the monolayer is removed without damaging the substrate for these fluences.

Figure 2(d) displays Raman spectra taken across the hole in Figure 2(a), with each spectrum displaced vertically for clarity. The color-coded circles in Figure 2(b) mark the positions where the Raman signals were collected. The common features of a broad peak at ~ 1180 cm$^{-1}$ and a smaller one at 1553 cm$^{-1}$ are attributed to the quartz substrate (see section S2). The $E_{2g}$ in-plane optical phonon Raman peak at 1367 cm$^{-1}$ is weak yet clearly resolved and consistent with literature values of 1366 – 1370 cm$^{-1}$ for hBN monolayers [1, 23]. This peak vanishes within the hole (traces #5-7), indicating a clean removal of the film by the ablation process. The Raman spectra near the border of the hole (traces # 3,4, 8, 9) show no new peaks or frequency shifts of existing peaks within our detection limit. The former is in sharp contrast to a previous report [8], in which a new Raman peak at 1295 cm$^{-1}$ corresponding to cubic boron nitride (cBN) nano-crystals was observed when the ablation was incurred by 80-MHz nano-joule femtosecond pulses. The latter, if they exist, would indicate the presence of strains originating from vacancies and lattice disorder introduced by the ablation, which was observed in MoS$_2$ monolayers [24]. The lack of such changes reveals that hBN monolayers remain pristine around the edge of the ablated hole, even at an intensity close to the ablation threshold. This has important implications for practical applications.

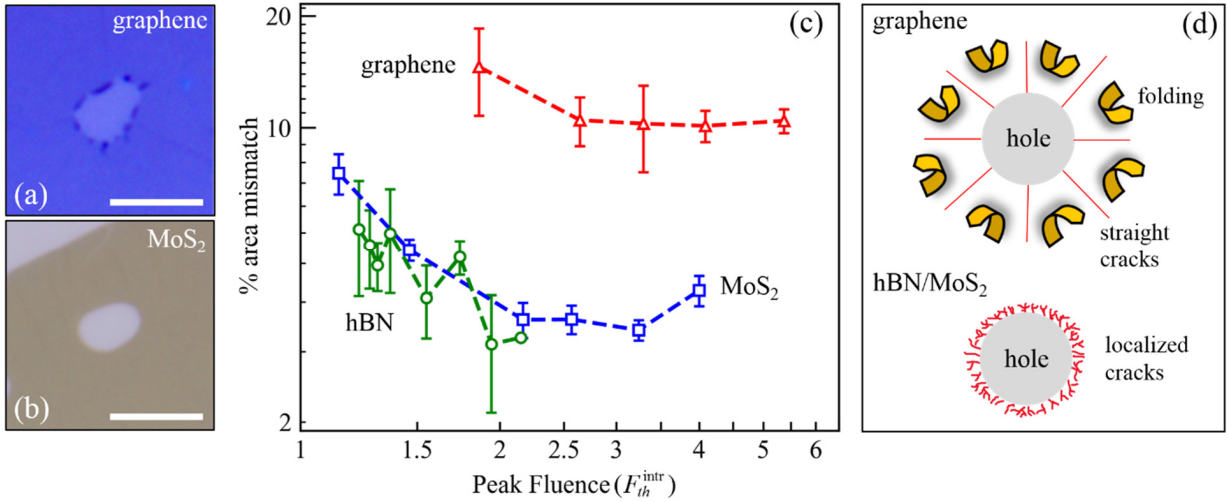

Figure 3: (a) An ablated hole in graphene film displaying folding around the edges, and (b) an ablated hole in MoS$_2$ monolayer flakes displaying smooth edges. Both results were performed on 90 nm SiO$_2$-Si substrates with a laser beam waist of 2.2 µm. Both scale-bars are 5 µm. (c) Calculated percentage area mismatch vs peak fluence (in units of $F_{th}^{intr}$) data for hBN, MoS$_2$, and graphene. (d) Illustration (not to scale) of different crack propagation in hBN (MoS$_2$) and graphene.

Figure 2(a) reveals that the circumference of the ablated hole is smooth and matched to the laser beam profile. A similar result was obtained for MoS$_2$ monolayer (Figure 3(a)). This is in stark contrast to graphene, which exhibits petal-like folds outwards away from the ablated hole (Figure 3(b) or Ref. [25] for a demonstration with stronger contrast). Graphene folding was found universally on supporting substrates, possibly due to the rapid substrate expansion during the laser heating [25], which makes high-fidelity patterning of graphene difficult using femtosecond ablation [26]. To quantify the fidelity, the percentage area mismatch is calculated (see section S3) and plotted as a function of the peak fluence (in units of $F_{th}^{intr}$) for graphene, MoS$_2$, and hBN (Figure 3(c)). As shown, the percentage area mismatch is comparable for hBN and MoS$_2$ monolayers but is 4× smaller than that of graphene. We attribute this sharp contrast in fidelity to their different fracture toughness, which is the ability of a material to resist catastrophic fracture [27, 28]. Take graphene and hBN as examples. The removal of atoms in the film via laser ablation introduces compressive stress on the edge of the film [29] and cracks initiate from stress concentration sites. For graphene, the cleavage of the bonds results in two identical zigzag edges. Because of this symmetric edge stress, the crack tip propagates along a straight line over a longer distance (*i.e.*, low fracture toughness) [30]. Upon the fast substrate expansion induced by the laser heating, the remaining graphene between two adjacent cracks fold [25], leading to the higher area mismatch and lower fidelity (Figure 3(d)). For hBN, the situation is entirely different [28]. The cleavage of h-BN bonds results in two different types of zigzag edge: a B-terminated edge (B-edge) and an N-terminated edge (N-edge). Because of this asymmetric edge stress, the crack tip bifurcates first and then deflects from its original propagation direction, and consequently the B-edge and N-edge swap their positions relative to the crack tip owing to their three-fold symmetry [28]. Cracks propagate via repeated deflection and sometimes branch, dissipating substantial energy to form a high density of localized damage close to the edge of the hole (*i.e.*, high fracture toughness). Experiments have shown hBN has a fracture toughness up to 10× higher than graphene [28, 30]. The same fracture physics can be applied to other heterogeneous 2D crystals with 3 fold symmetry such as $MoS_2$ [28], which is supported by our data in Figure 3(c). Our result indicates femtosecond laser patterning of hBN and $MoS_2$ has high fidelity and can generate deterministic features.

## Bandgap scaling of the ablation threshold for 2D materials

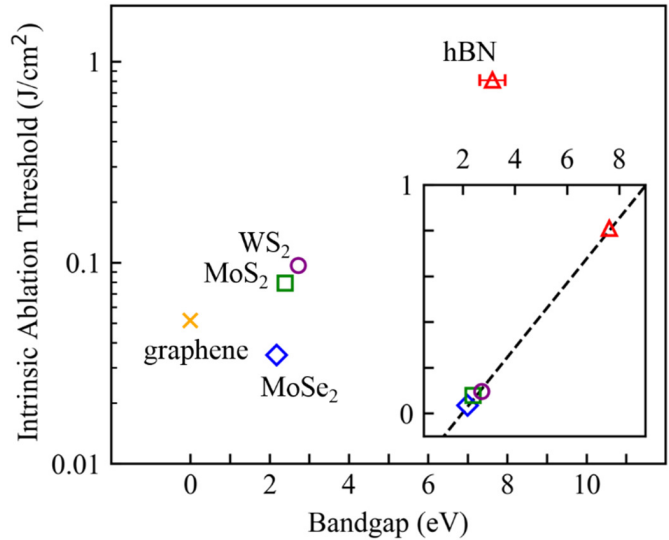

Figure 4: (color online) A log-linear plot of $F_{th}^{intr}$ vs. electronic bandgap of monolayer 2D materials. The inset shows the same plot in linear-linear scale excluding graphene. $F_{th}^{intr}$ are 51 mJ/cm$^2$ (graphene, blue), 35 mJ/cm$^2$ ($MoSe_2$, yellow), 79 mJ/cm$^2$ ($MoS_2$, green), 97 mJ/cm$^2$ ($WS_2$, purple), 800 mJ/cm$^2$ (hBN, red).

Breakdown experiments like Figure 1 were carried out for other monolayer 2D materials. Figure 4 shows the dependence of the $F_{th}^{intr}$ on the quasi-particle bandgap energy for graphene, $MoSe_2$, $MoS_2$, $WS_2$, and hBN monolayers. Although $MoSe_2$ has a bandgap energy ($E_g \sim 2.18$ eV) larger than the photon energy of 1.55 eV, it has a strong excitonic resonance at 1.57 eV [31]. Both graphene and $MoSe_2$ therefore exhibit strong saturable linear absorption at 800 nm, which is why their $F_{th}^{intr}$ are among the lowest. Despite its zero bandgap, graphene has a higher $F_{th}^{intr}$ than $MoSe_2$, which we attribute to its exceptional mechanical strength, including 10× larger Young's modulus and breaking strength (see section S5). On the other hand, carrier generation in MoS$_2$ ($E_g$ ~ 2.40 eV), WS$_2$ (~ 2.73 eV)[31] and hBN (7.6 ± 0.3 eV) monolayers, which have bandgap energy higher than the photon energy, is expected to be initiated by nonlinear ionization followed by avalanche ionization. The Keldysh parameters for these materials at their respective breakdown thresholds are estimated to be 2.16, 2.11, and 1.57, respectively (see section S6), indicating that the nonlinear ionization is in the intermediate regime between the multiphoton and tunneling [32]. The inset in Figure 4, excluding graphene, reveals a linear relationship between $F_{th}^{intr}$ on the bandgap energy. Similar trends were reported for bulk materials, although there is a lack of consensus in understanding. Joglekar et al. studied dielectric breakdown in Si, fused SiO$_2$, quartz and Al$_2$O$_3$ with 1053-nm, 800-fs pulses [20]. They reported a linear dependance of breakdown threshold on the bandgap energy, which was attributed to the bandgap scaling of the pondermotive energy (linearly proportional to the laser intensity) in avalanche ionization. Mero et al. studied dielectric breakdown in sputtered thin films of TiO$_2$, Ta$_2$O$_5$, HfO$_2$, Al$_2$O$_3$, and SiO$_2$ with 800-nm pulses between 25 fs to 1.3 ps [33]. They also observed a linear dependance, which was attributed to the bandgap dependence of the multiphoton absorption from Keldysh photoionization theory. A similar result was also reported for SF11, Corning 0211 glass, fused silica, and CaF$_2$ with both 800- and 400-nm femtosecond pulses [34]. The linear dependency suggested in the inset of Figure 4, if substantiated by more data points from other 2D materials with a bandgap energy between 3 and 7 eV, would clearly validate a universal scaling law that is independent of materials' dimensinality. To understand the physics of this linear dependancy, systematic studies need to be carried out to differentiate between the roles of nonlinear and avalanche ionization in 2D materials. This scaling law is practically very useful in predicting $F_{th}^{intr}$ for other 2D materials.

2D vs. 3D dielectric breakdown

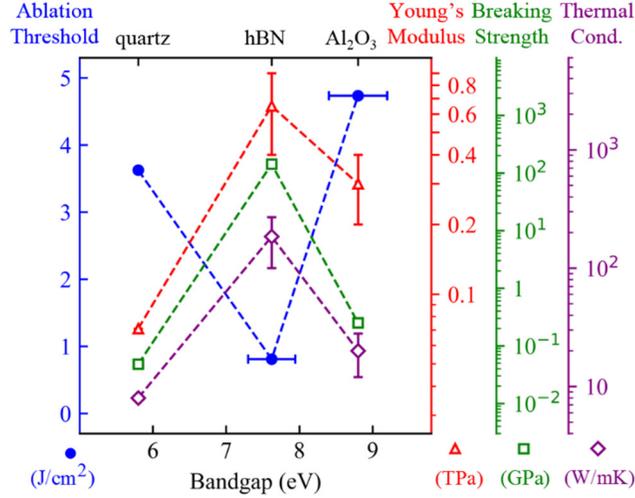

Figure 5: Comparison of the ablation threshold, Young's modulus, breaking strength, and thermal conductivity of fused quartz, hBN monolayers, and $Al_2O_3$. Their bandgap energies are ~ 5.8 eV, 7.6 ± 0.3 eV, and 8.8 ± 0.4 eV, respectively.

It is interesting to point out that, although the bandgap of hBN monolayers (7.6 ± 0.3 eV) is higher than that of fused quartz (~ 5.8 eV), the latter remains intact at the breakdown threshold of hBN (Figure 2(a)). This is counterintuitive as materials with large bandgaps are expected to have higher breakdown thresholds. This is further illustrated in Figure 5, which displays the surface breakdown threshold $F_{th}$ for fused quartz (3.6 J/cm$^2$) and $Al_2O_3$ (4.7 J/cm$^2$) (see section S7), and the intrinsic breakdown threshold $F_{th}^{intr}$ of hBN (0.8 J/cm$^2$). Even though hBN monolayers have a bandgap energy between quartz and $Al_2O_3$ ($E_g$ ~ 8.8 ± 0.4 eV), its $F_{th}^{intr}$ is 4× smaller than that of quartz and 6× smaller than that of $Al_2O_3$. This is even more surprising if we consider that hBN monolayers have a Young's modulus comparable to $Al_2O_3$ and is ~10× larger than quartz, a breaking strength 550× larger than $Al_2O_3$ and 3000× larger than quartz, and a thermal conductivity ~10× larger than $Al_2O_3$ and 10-30× larger than quartz (see right vertical axes of Figure 5; also see Section S5). Given hBN is more robust mechanically, dissipates energy faster, and yet is easier to ablate, our data strongly suggest that the carrier generation in hBN is much more efficient than in quartz and $Al_2O_3$. At breakdown, the Keldysh parameters [32] for quartz, hBN, and $Al_2O_3$ are estimated to be ~0.83, ~1.57, and ~0.80, respectively, which again indicates the nonlinear ionization is in the intermediate regime between the multiphoton and tunneling (see section S6). In the literature, it was reported that $MoS_2$ monolayers have 2- and 3-photon absorption coefficients $10^3$× larger than typical bulk materials [35, 36]. The authors attributed this enhancement to excitonic effect, even though the detuning is substantial between the exciton resonance and the photon energy used in those experiments. Such a process can be expected for hBN as well, especially it has a true exciton resonance at 5.6 - 6.3 eV (corresponding to 4-photon absorption of 800 nm) [18], although this has not been experimentally demonstrated. We hypothesize the enhanced electronic density of states inherited in the lower dimensional systems near the band edge could also contribute to enhanced nonlinear absorption [37]. On the other hand, Auger recombination rates were reported to be $10\times$ - $10^6\times$ larger in transition metal dichalcogenide (TMD) monolayers than the bulk materials, which were attributed to enhanced Coulomb interaction as a result of quantum confinement, reduced dielectric screening, and the large effective masses of electrons and holes in TMDs [38, 39]. This implies that avalanche ionization, which is regarded as the reverse process of Auger recombination, is expected to be enhanced as well in 2D monolayers, although it has not been experimentally demonstrated, either. Future experiments need to be designed and carried out to verify these conjectures. This finding is useful in that it enables materials with lower bandgaps, rather than only transparent substrates with extremely large bandgaps ($\geq$ 8 eV), to support or interface with hBN monolayers for strong-field study. For example, diamond has a relatively low bandgap (~ 5.5 eV) and $10\times$ better thermal conductivity than hBN. In the quest for high harmonic generation in hBN, it may thus be advantageous to use diamond substrates to dissipate heat efficiently and avoid thermal damage under repetitive high intensity excitation.

## Sub-micron resolution patterning

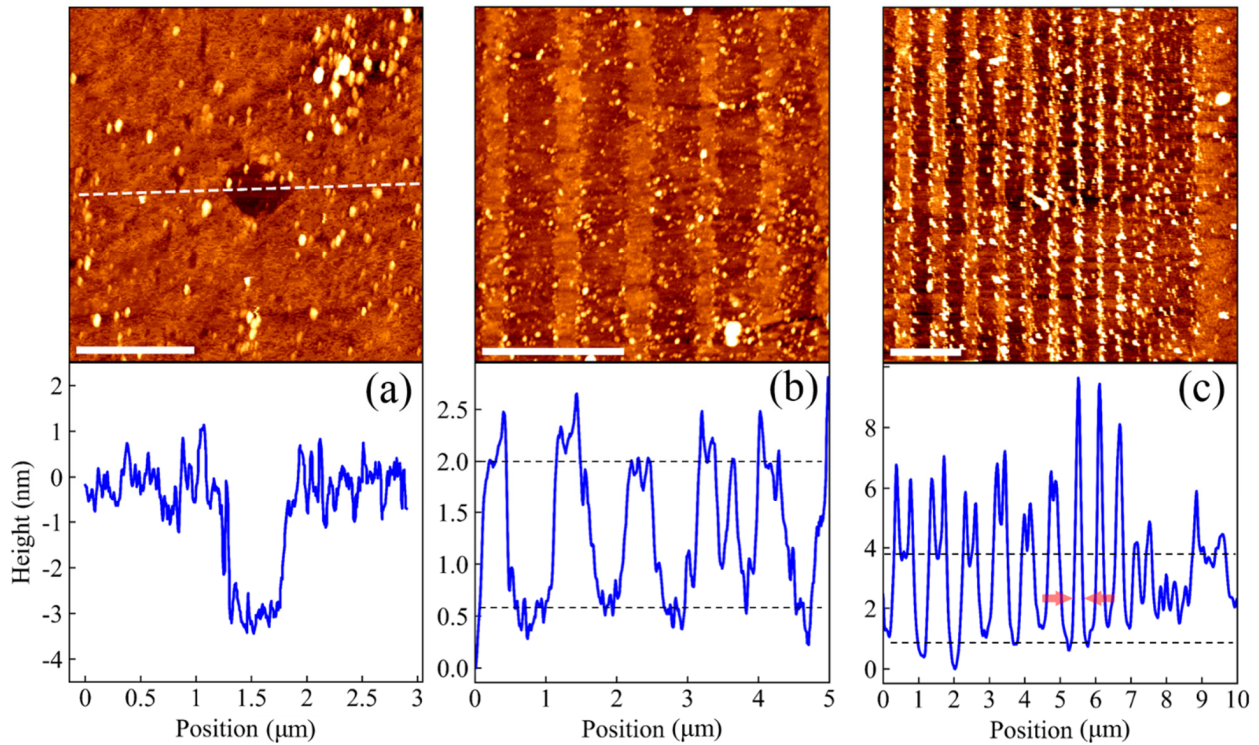

Figure 6: Laser patterning of hBN monolayer on quartz. AFM image (top) and column-averaged cross-sectional profile (bottom) for (a) an isolated hole obtained with 1.1 $F_{th}^{intr}$, (b) periodic stripes obtained with 2 $F_{th}^{intr}$, and (c) an array of stripes with decreasing width obtained with 1.5 $F_{th}^{intr}$. The dashed lines mark the top and bottom of the film. The white particles are PMMA residues. The scale bars are 1, 2, 2 µm for a, b, c, respectively.

By using a high-NA objective and a fluence close to the breakdown threshold, we demonstrated ablated features in hBN monolayers with sub-wavelength resolution. Figure 6(a) shows an AFM image of an isolated hole in hBN on quartz ablated with a single pulse focused by 0.9 NA objective at 1.1 $F_{th}^{intr}$. The cross-sectional profile at the bottom indicates a hole diameter of ~500 nm. Figure 6(b) shows an AFM image of an array of stripes obtained by ablating periodic lines in the hBN monolayer using a 100-kHz laser beam with a fluence ~ 2 $F_{th}^{intr}$ and a sample translation speed of 100 µm/s. The bottom plot displays a column-averaged cross-sectional profile obtained by stacking all columns and then dividing it by the number of columns to remove low-frequency spatial noise. The offset of the laser beam is 1 µm, producing a stripe width ~ 400 nm. The stripes display width variation and edge roughness due to the finite positioning accuracy of the translation stage in the depth and lateral directions, which was hampered by the cross coupling from the moving vertical axis. This cross coupling is especially evident when using a high NA objective with a Rayleigh range of less than 500 nm that was determined experimentally. Figure 6(c) shows an AFM image and its column-averaged profile of an array of stripes obtained with a fluence of ~ 1.5 $F_{th}^{intr}$. The laser beam offset starts at 1 µm and decreases by 50 nm per line, leading to a gradual decrease of the strip width. The PMMA residues are more pronounced in this region of the sample, especially along the edges of the stripes, which suggests these PMMA particles may have aggregated upon laser irradiation. These aggregates lead to the fork-like features in the row-averaged height profile, which evolve into peaks for very small lateral offsets of the laser beam. The horizontal dashed lines delineate the top and bottom of the film, from which the narrower stripe width is determined to be ~ 200 nm. When the beam offset is too small, hBN stripes become partially removed due to positioning instability. Currently, stripe width and edge smoothness are stage limited; narrower stripes than what is demonstrated here can be expected with a more stable translational stage or employing beam shaping techniques [14]. For future experiments, a more effective recipe for cleaning PMMA residue [11] or a residue-free transferring agent such as PDMS [40] can be adopted.

## Conclusions

In conclusion, we report the first systematic study of dielectric breakdown of hBN monolayers induced by single femtosecond laser pulse. The intrinsic breakdown threshold fluence $F_{th}^{intr}$ was determined to be 0.8 J/cm$^2$, the highest among all two-dimensional materials. Enabled by the much larger bandgap of hBN compared to TMDs, we observe for the first time a clear linear dependence of breakdown threshold on the bandgap energy in 2D materials, demonstrating that such a linear dependency is a dimensionally independent scaling law. This finding is useful for predicting the breakdown thresholds of other 2D materials. Moreover, our data reveals that, although hBN monolayers have a higher bandgap energy and mechanical strength than fused quartz, it is optically weaker (4× lower in breakdown threshold). This counterintuitive observation implies that carrier generation processes, including both nonlinear and avalanche ionizations, are much more efficient in hBN monolayers compared to quartz. This finding opens up more choices of supporting substrates to study strong-field optical phenomena in hBN. On the practical side, we show femtosecond laser ablation removes hBN monolayers without damaging the underlying substrates or the surrounding hBN film. The ablated features in hBN and MoS$_2$ monolayers were found to possess relatively small edge roughness compared to graphene, which we attribute to their high fracture toughness due to asymmetric edge stress. Finally, we demonstrate femtosecond laser patterning of isolated holes of ~500 nm diameter and strips of ~200 nm width in hBN monolayers. Our work pushes the frontier of strong-field phenomena of hBN and firmly establish femtosecond laser as a novel and promising tool for one-step deterministic patterning of hBN monolayers.

## Acknowledgements

We greatly appreciate Dr. Jimmie A. Miller (UNCC) for assistance in AFM measurements. We also thank Professor Stuart T. Smith (UNCC) for discussion on fracture toughness, and Professor Haitao Zhang and Mr. Simon M. Sami (UNCC) for use of the Shimadzu spectrophotometer and hot plates. We also thank Mr. Kenan Darden (UNCC) for proofreading. W.-H.C. acknowledges the support from the National Science and Technology Council of Taiwan (NSTC 111-2119-M-A49-005-MBK).

# Dielectric breakdown and sub-wavelength patterning of monolayer hexagonal boron nitride using femtosecond pulses

Sabeeh Irfan Ahmad[1], Emmanuel Sarpong[1], Arpit Dave[1], Hsin-Yu Yao[2], Joel M. Solomon[1], Jing-Kai Jiang[3], Chih-Wei Luo[3], Wen-Hao Chang,[3,4] and Tsing-Hua Her[1]

[1]Department of Physics and Optical Science, The University of North Carolina at Charlotte, 9201 University City Boulevard, Charlotte NC 28223
[2]Department of Physics, National Chung Cheng University, No.168, Sec. 1, University Rd., Minhsiung, Chiayi 621301, Taiwan.
[3]Department of Electrophysics, National Yang Ming Chiao Tung University, Hsinchu 30010, Taiwan
[4]Research Center for Applied Sciences, Academia Sinica, Taipei 11529, Taiwan

Email: ther@uncc.edu,

## Supplementary Information

## S1. Bandgap of supporting substrates

The bandgaps of $Al_2O_3$ and quartz supporting substrates were determined from their UV-VIS absorbance spectra between 185 and 500 nm, using a Shimadzu UV2600 spectrophotometer. The data is displayed in Figure S1. The spectrum for $Al_2O_3$ exhibits minimal absorption with a tail beginning ~ 6 eV, which leaves its bandgap undetermined. A more significant signal is observed for quartz. Its bandgap $E_g$ is estimated to be ~ 5.8 eV, by fitting the measured absorbance data to the equation $(\alpha l)^2 \propto (h\nu - E_g)$ for direct bandgap materials [1], where $\alpha$ is the absorption coefficient, $l$ is the sample thickness, $h$ is planck's constant, and $\nu$ is the frequency.

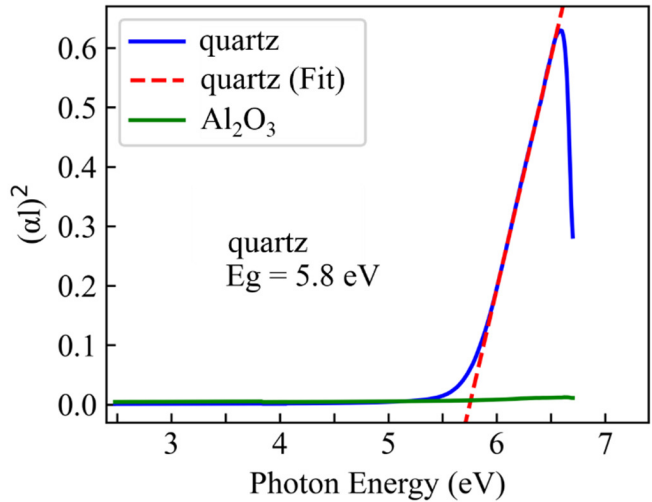

Figure S1: UV-VIS absorbance squared vs. photon energy (eV). The bandgap is estimated to be ~ 5.8 eV for quartz.

## S2. Raman signals of supporting substrates

The Raman spectra for the bare $Al_2O_3$ and quartz substrates are displayed in Figure S2, showing the necessity of doing experiments on quartz if Raman signals are to be collected. $Al_2O_3$ exhibits a peak at ~1356 cm$^{-1}$, which interferes with that of hBN monolayers at ~1367 cm$^{-1}$.

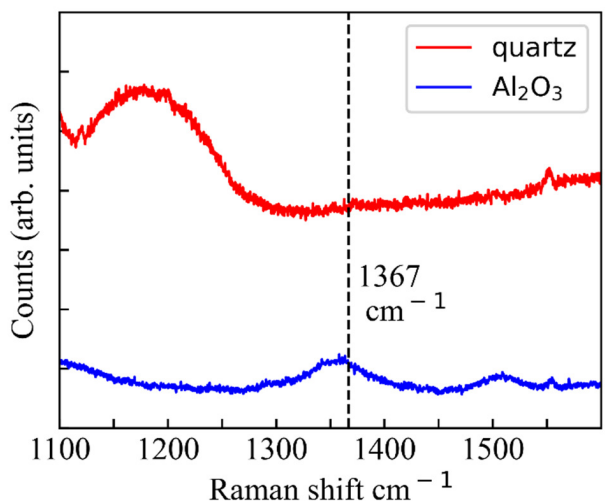

Figure S2: Raman spectra for quartz and $Al_2O_3$. The spectra are shifted vertically for clarity. The $Al_2O_3$ spectrum shows a peak near 1356 cm$^{-1}$ that obscures hBN's Raman peak at 1367 cm$^{-1}$.

## S3. Fidelity analysis of femtosecond laser patterning

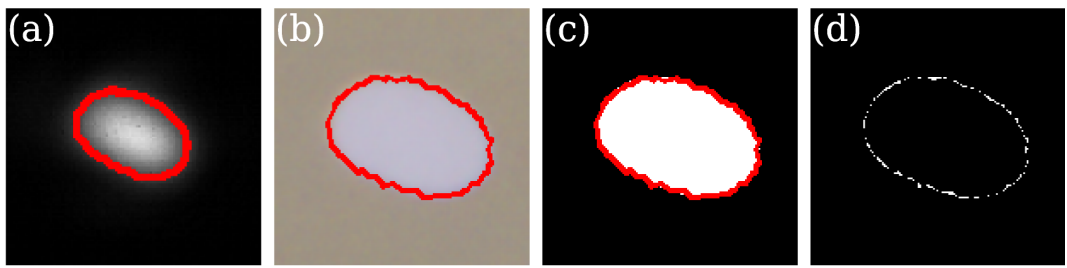

Figure S3: Fidelity analysis of laser patterning. (a) The spatial profile of the laser spot at focus. The red outline represents the level of constant intensity based on a laser profile intensity level or clip level. (b) Ablation hole on a $MoS_2$ monolayer with the spatial profile of the laser (from (a)) matched to the ablation hole. (c) A thresholded image of the ablation hole (white area means ablated region) with the matched laser profile. (d) A difference image whose pixels do not overlap either the fitted laser spatial profile or the ablation profile.

To quantify the regularity of the hole, the spatial profile of the beam was matched to the ablation holes in the monolayers, as outline in Figure S3 which uses $MoS_2$ as an example. By positioning, rotating, scaling, and setting the laser profile intensity level or clip level, the laser spot can be fitted to the ablation area to measure how well the laser spot matches the ablation features (Figures S3(a)-(c)). To quantify this fit, the difference between the fitted laser spot 2D profile for a chosen clip level and a thresholded ablation 2D profile were taken, resulting in an image whose pixels do not overlap either the laser spatial profile or the ablation profile as shown in Figures S3(d). The pixel difference was then normalized to the number of pixels that make up the ablation feature to become the percentage area difference, as defined in the main text. The best clip level is chosen to minimize the percentage area difference.

## S4. Surface roughness of the ablated hole and the bare substrate

Figure S4 compares histograms of AFM height data within the ablated hole of Figure 2(a) on quartz and of the bare quartz substrate. Both histograms contain the same number of pixels. The widths of these two histograms are similar, indicating the monolayer is removed without damaging the substrate.

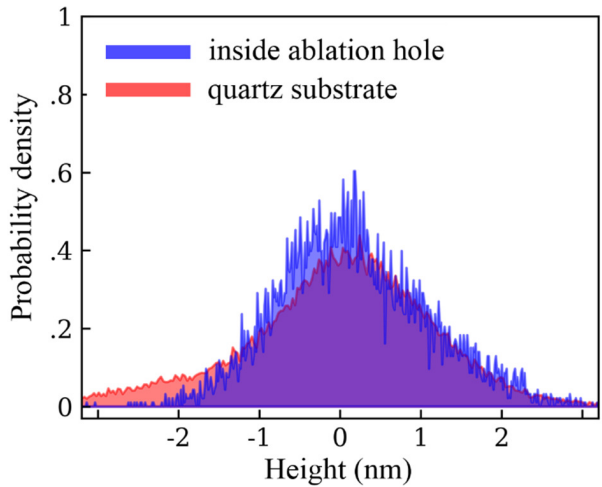

Figure S4: Histograms of AFM height data of the ablation hole (blue trace) and of the bare quartz substrate (red trace). The vertical axis is the probability density function (PDF).

## S5. Mechanical and thermal properties of materials

**Error! Reference source not found.**Table S1 displays the mechanical and thermal properties of various bulk and 2D materials studied in this work.

Table S1: Mechanical properties of 2D (graphene and hBN) and bulk ($Al_2O_3$ and quartz) materials from literature sources.

|  | Young's Modulus (TPa) | Breaking Strength (GPa) | Thermal Conductivity (W/mk) | Refs |
|---|---|---|---|---|
| graphene | 1 | 130 | 600 | [2, 3] |
| MoSe$_2$ | 0.1 | 13 | 44 | [4, 5] |
| hBN | 0.4-0.9 | 120-165 | 100-270 | [2] |
| Al$_2$O$_3$ | 0.2-0.4 | 0.25 | 12-28 | [6] |
| quartz | 0.071 | 0.048 | 8 | [7] |

## S6. Keldysh parameters

Table S2: Material parameters for the calculation of the Keldysh parameters for 2D and bulk materials. Laser beam parameters are $\lambda$ = 800 nm and $\tau$ = 160 fs.

| | Electronic Bandgap (eV) | $m_e$ | $m_h$ | $F_{th}$ (J/cm$^2$) | Keldysh Parameter | Refs |
|---|---|---|---|---|---|---|
| MoS$_2$ | 2.40 | 0.43 | 0.43 | 0.079 | 2.16 | [8, 9] |
| WS$_2$ | 2.73 | 0.44 | 0.45 | 0.097 | 2.11 | [8, 9] |
| hBN | 7.2 - 8.2 | 0.83 | 0.63 | 0.8 | 1.57 | [10, 11] |
| quartz | 5.8 | 0.39 | 7.5 | 3.6 | 0.83 | Section S1,[12] |
| Al$_2$O$_3$ | 8 - 9.4 | 0.39 | 6.2 | 4.7 | 0.80 | [13-15] |

Keldysh parameter [16], defined by $\gamma = \omega\sqrt{mE_g}/(eE)$ where $\omega$ is the light frequency, $m$ is the reduced mass of the electron-hole pair, and $E$ is the electric field, is calculated for various bulk and 2D materials at their respective breakdown threshold, as shown in Table S2.

## S7. Surface ablation threshold of supporting substrates

Figure S5 shows the Liu plot for the bare Al$_2$O$_3$ and quartz substrates. The bean radius $\omega_o$ and surface ablation threshold $F_{th}$ parameters are determined to be respectively 4.2 µm and 3.63 J/cm$^2$ for quartz and are 3.8 µm and 4.7 J/cm$^2$ for Al$_2$O$_3$ respectively.

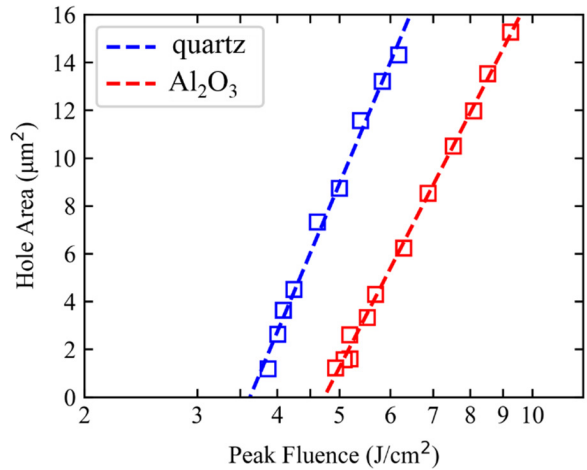

Figure S5: Hole area vs peak fluence for bare Al$_2$O$_3$ and quartz substrates.